\documentclass[intlimits,twoside,a4paper]{article}
\usepackage{wrapfig}
  \usepackage{amsmath,amssymb} 
  \usepackage{graphicx}        

\usepackage[cp1251]{inputenc}
\usepackage{xcolor}
\usepackage[eqsecnum]{cmpj3}

\usepackage{bm}
\issue{2020}{23}{2}{23606}
\doinumber{10.5488/CMP.23.23606}

\title[Liquid 3$d$ transition metals]%
{First principles determination of some static and dynamic 
properties of the liquid 3$d$ transition metals near melting%
}
\author[B.G. del Rio {\it et al.}]{B.G. del Rio, C. Pascual, O. Rodriguez, L.E. Gonz\'alez,
D.J. Gonz\'alez}
\address{
 Departamento de F\'\i sica Te\'orica,
Universidad de Valladolid, 47011 Valladolid, Spain
}
\date{Received January 31, 2020, in final form March 26, 2020}

\begin{document}

\maketitle

\begin{abstract}

We report an \textit{ab initio} molecular dynamics simulation 
	study of several static and dynamic properties of the 
liquid 3\textit{d} transition  metals. 
The calculated static structure factors show  qualitative agreement with 
the available experimental data, and  its second peak displays 
an asymmetric shape which suggests  a significant   
local icosahedral short-range order. 
The dynamical structure reveals propagating density fluctuations 
whose dispersion relation has been evaluated; moreover, its long wavelength 
	limit is compatible with their respective experimental sound velocity. 
Results are reported for the longitudinal and 
transverse current spectral functions as well as for 
the respective dispersion relations. 
We also analyze the possible appearance of transverse-like low-energy excitations in the 
calculated dynamic structure factors.
	Several transport coefficients have been evaluated and compared with the 
	available experimental data. 

\end{abstract}

\section{Introduction}

This paper presents  recent 
{\it ab initio} molecular dynamics simulation studies  
for a range of bulk  static, dynamic and electronic properties of the  
liquid 3$d$ transition  metals at thermodynamic conditions near melting. 
We review some of our previous calculations 
\cite{RRGGTi,MGGFe,RGGAgNi,RG17,RPGGlam}   
along with several new results.    

Despite the practical relevance of 
the 3$d$ transition  metals,  it is 
remarkable that
the experimental information concerning their
static and dynamic properties in the liquid state is 
rather scarce. 

The static structure factor, $S(q)$, encompasses the information
about the average short-range order of a liquid system, 
and was first obtained experimentally by 
Waseda and coworkers \cite{WasedaBook} in the decade of the 1970s
for the 3$d$ transition metals, 
by means of 
X-ray diffraction (XD) techniques.  More recently,  
the $S(q)$ of l-Ti, l-Fe, l-Ni and l-Zn have been determined again 
by either XD or neutron diffraction (ND) experiments \cite{SqFeNi,SqFeNi2,SqTi,SqTi2,SqZn}.
In l-Zn, the new measurements showed a good agreement with the older ones, but 
for l-Ni and l-Fe, 
the comparison with Waseda and coworkers' XD (WXD) data for the $S(q)$, revealed  
some discrepancies concerning the height of the
main peak as well as the shape of the second peak; nevertheless, 
the positions of the maxima and minima 
of $S(q)$ were coincident in all the experiments. 
On the other hand, the  
recent XD and ND data for the $S(q)$ of l-Ti \cite{SqTi,SqTi2} yielded 
results well compatible between each other, but 
when compared with WXD data, they were displaced towards 
greater $q$-values by a significant amount of $\approx$ 0.20 \AA$^{-1}$. 
Moreover, the recent XD data showed 
a second peak with a marked shoulder on the high-$q$ side; actually,   
shoulders on the high-$q$ side of the second peak 
of $S(q)$ have also been found experimentally in other transition 
metals (Fe, Ni, Zr) \cite{SqFeNi,SqFeNi2,Kelton-TiZrNi}. 
We think that these disparities call for  
new precise structural data for the other $3d$ transition metals, both
extracted from new additional experiments and/or from accurate simulations 
as the {\it ab initio} ones presented here.

Several transport coefficients of most of the $3d$ liquid metals were reported in the 
literature~\cite{Blairs,Casas,Assael,Egry,SaidHorbatch,Agaev,IidaBook,Nash,ShimoBook,Hirai,Nachtrieb,Assael2,Marcus}.  
The dynamic structure factors of l-Ti, l-Fe, 
l-Ni and l-Zn near melting 
were measured by 
both inelastic neutron (INS) and 
inelastic X-ray (IXS) scattering 
\cite{SaidHorbatch,Hosokawa,JohnBerCha,JohnBerCha1,HosoZana,HosoZana1}, whereas 
the self dynamic structure factors of l-Ni, l-Ti and l-Cu were determined by 
quasi-elastic neutron scattering \cite{Chato,Horbach,Meyer}.  
There is, however, no experimental information about the dynamic structure of
the rest of $3d$ liquid metals.

To our knowlege, only l-Cu, l-Ni and l-Fe had been previously studied by 
other groups using {\it ab initio} simulation methods. 
For l-Cu,  three {\it ab initio} studies evaluated its static properties 
at melting and some undercooled states  \cite{Pasqua,Pasqua1,JakPas}.  
As for l-Ni,  
Jakse {et al.} \cite{Jakse2004,Jakse2007} 
studied a range of static and dynamic
 properties near melting and calculated the self-diffusion 
and shear viscosity  coefficients. In the case of l-Fe, Ganesh and Widom
\cite{Ganesh} analyzed several static properties, getting also into
the undercooled region.

In our recent studies of the liquid 3$d$ transition  metals 
\cite{RRGGTi,MGGFe,RGGAgNi,RG17,RPGGlam}, we were interested in 
the calculation and analysis of their  
dynamic properties. Besides its intrinsic interest, we were also motivated  by 
the recent finding of some remarkable features 
such as low-energy transverse-like excitations in the dynamic structure factor and/or the 
appearance of a high-frequency branch in the transverse current dispersion 
relation, in addition to the usual one that appears at low frequencies. 
Indeed, the study of the microscopic processes behind the 
dynamic properties of liquid systems has
been a long standing research field of Prof. Mryglod, and in this respect we spotlight his 
contributions to the development and application of the generalized collective modes (GCM) 
approach. 
As for the above mentioned features, there is still no clear explanation of their physical origin, 
although we have recently advanced some connection between the existence of a 
second branch in the transverse dispersion relation and the coupling of the transverse 
current with density fluctuations at all wavevectors \cite{RG17,RCGCSn18}.   

Our studies of the bulk liquid 
3$d$ transition  metals were performed 
with an {\it ab initio} molecular dynamics (AIMD) simulation method
based on the density functional theory \cite{HK,KS}. The liquid 
metal is modelled as an interacting system of ions and electrons where  
the ionic positions evolve according to
classical mechanics while the electronic subsystem follows adiabatically.
Table~\ref{states} provides information concerning the 
thermodynamic states as well as some technical details.  The  
AIMD simulations were performed with the 
Quantum-ESPRESSO package \cite{espresso,espresso1}, excepting l-Fe and l-Zn
for which we used the VASP code \cite{vasp,vasp1,vasp2,vasp3}.   
The generalized gradient approximation of Perdew-Burke-Ernzerhof~\cite{PBE} was used to account for  
the electronic exchange-correlation 
energy, with the exceptions of  
l-Ti and l-Cu for which we used 
Perdew and Wang's approximation \cite{PW} 
and Perdew and Zunger's local density 
expression~\cite{PZ}, respectively.

\begin{table}[!b]
\vspace{-3mm}
\caption{
Input data of the liquid 3$d$ transition metals considered in 
the present AIMD simulation study.  $\rho$ is the total ionic number
density, (taken from \cite{BBGSS1}), $T$ is the temperature, 
$N_{\rm part}$ is the number of particles, 
 $\Delta t$ is the ionic timestep, $Z_{\rm val}$ is the number of 
valence electrons  and 
$N_\text{c}$ is the total number of configurations. 
\label{states}
}
\vspace{2ex}
\begin{center}
\begin{tabular}{cccccccc}
\hline
& $\;\;\;$  &  $\rho$ \big(\AA$^{-3\strut}$\big)   $\;\;\;$ & $T$ (K) & $N_{\rm part}$ 
& $\Delta t$ (ps)  & $Z_{\rm val}$ & $N_\text{c}$  \\
\hline
& Sc & 0.0391 &  1875  &120 & 0.0045 & 11 & 29000  \\
& Ti & 0.0522  &  2000  &100& 0.0045 & 12 & 30700   \\
& V & 0.0634  &  2173  & 120 & 0.0040 & 13 & 20000  \\
& Cr & 0.0713  &  2173  & 100 & 0.0045 & 14 & 27000  \\
& Mn & 0.0655  & 1550  & 120 & 0.0045 & 15 & 13200  \\
& Fe & 0.0746  & 1873  & 100 & 0.0020 & 8 & 22100  \\
& Co & 0.0787  & 1850  & 135 & 0.0045 & 9 & 13100  \\
& Ni & 0.0792  & 1773  & 120 & 0 0055 & 10 & 21000  \\
& Cu & 0.0755  & 1423  & 150 & 0 0050 & 11 & 25500  \\
& Zn & 0.0616  &  723  &  90 & 0 0040 & 12 & 36000  \\
\hline
\end{tabular}
\end{center}
\end{table}


The ion-electron interaction was described by  
ultrasoft pseudopotentials \cite{vanderbilt},  
excepting l-Fe and l-Zn for which  the projector augmented
wave all-electron description \cite{PAW,PAW1} was used.  For all the systems,  
the cutoff for the plane-wave representation of the orbitals was within 
the range $25.0$--$35.0$ Ryd and 
the sampling of the Brillouin zone was performed by means of  
the single $\Gamma$ point.

We recall that most previous studies of the liquid 3$d$ transition metals 
were carried out  by using semiempirical or more fundamental pair (or many-body) potentials
coupled with liquid state theories or classical
molecular dynamics simulations \cite{Todd,HKH,BBGSS,BBGSS1}. Moreover, in 
these approaches the electronic degrees of freedom are
hidden into the effective potential and, therefore, no electronic 
properties (i.e.,  density
of states, conductivity, \ldots) can be consistently/reliably calculated.  
On the contrary, the  AIMD simulation method yields parameter-free interatomic 
interactions
derived from first principles, and the valence electrons are explicitly 
taken into account during the calculation.

\section{Results and discussion}
\label{results1}

The evaluation of static and dynamic  properties of the bulk 
liquid metal was 
performed by using the total number of 
equilibrium configurations listed in table \ref{states}.

\subsection{Static properties}
\label{results}

\begin{figure}[!b]
\centerline{\includegraphics[width=0.97\textwidth,clip]{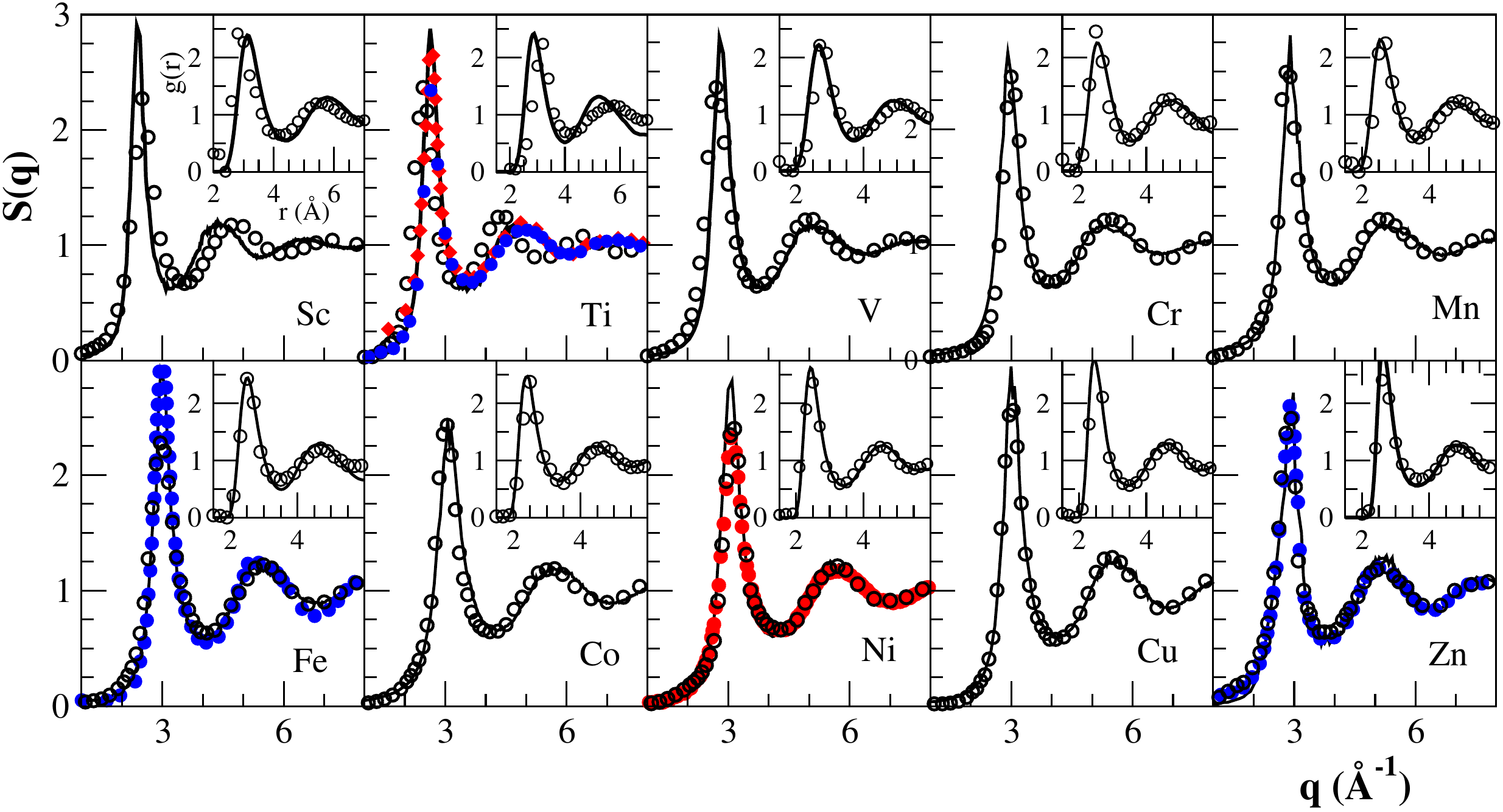}}
\caption{(Colour online) Static structure factor, $S(q)$, of the liquid 3$d$ transition 
metals. 
Continuous line: present AIMD calculations.
Open circles: WXD data
from \protect\cite{WasedaBook}.
The blue circles are XD data \cite{SqTi,SqTi2,SqFeNi,SqFeNi2,SqZn}
and red ones are ND data \cite{SqTi,SqTi2,SqFeNi,SqFeNi2}. 
The inset shows the corresponding pair distribution   
function.
\label{fig_sqA}
}
\end{figure}

The calculated static structure factors, $S(q)$, are  
plotted in figure \ref{fig_sqA} along with the 
corresponding WXD data  \cite{WasedaBook}. 
The figure also includes the 
more recent XD and ND  
data \cite{SqFeNi,SqFeNi2,SqTi,SqTi2,SqZn}  
for l-Ti, l-Fe, l-Ni and l-Zn. 

Aside from l-Sc and l-Ti, the calculated $S(q)$ shows a good qualitative 
agreement with the WXD data. There are some slight disparities, namely: 
(i) a very small phase shift in l-V, l-Mn and l-Co,   
(ii)~the calculations predict an asymmetric shape for the second peak with a 
 more or less marked shoulder, whereas the 
XD data display,  for all systems, a symmetric second peak.  
Notice that shoulders on the high-$q$ side of the second peak
of $S(q)$ have also been  experimentally observed in 
l-Ti, l-Fe, l-Ni and l-Zr~\cite{SqFeNi,SqFeNi2,SqTi,SqTi2,Kelton-TiZrNi}; moreover,
they become more marked upon undercooling \cite{KimKelton,Jakse,Jakse1}.

As for l-Ti, comparison with the 
 recent XD and ND data \cite{SqTi,SqTi2} for the $S(q)$, reveals an 
excellent agreement with 
the present resuts, including the shape 
of the second peak as well as the shoulder on its  high-$q$ side. 
Consequently, we believe that the 
WXD data for l-Ti are somewhat unreliable and that something similar 
might also happen with l-Sc, where an analogous (although opposite) dephasage is 
visible when compared with the present AIMD result.

From the calculated $S(q)$, within the range $q\leqslant 1.2$~\AA$^{-1}$, 
we used a least squares fit, $S(q)=s_0+s_2q^2$, 
 to obtain an estimate for $S(q \to 0)$ and the results are 
given in table \ref{static}. Then, 
we evaluated the 
isothermal compressibility, $\kappa_T$, by resorting to 
the relation 
$S(q \to 0)$ = $\rho k_{\text B} T \kappa_T$,  
where $k_{\text B}$ is Boltzmann's constant. 
Table \ref{static} lists the obtained results along with the available 
experimental data. 

In figure \ref{fig_sqA}, we  also depicted the 
associated 
pair distribution functions, $g(r)$, which are compared with 
their respective XD data \cite{WasedaBook}. 
The associated coordination number (CN) was evaluated by 
integrating the radial distribution function, $4\piup \rho r^2 g(r)$, up to the position of its
first     
minimum, $r_{\rm min}$. Table \ref{static} shows the obtained values 
for the CN's which are typical 
of simple liquid metals near their respective triple points~\cite{Balubook}.

\begin{wrapfigure}{r}{0.42\textwidth}
\centerline{\includegraphics[width=0.46\textwidth,clip]{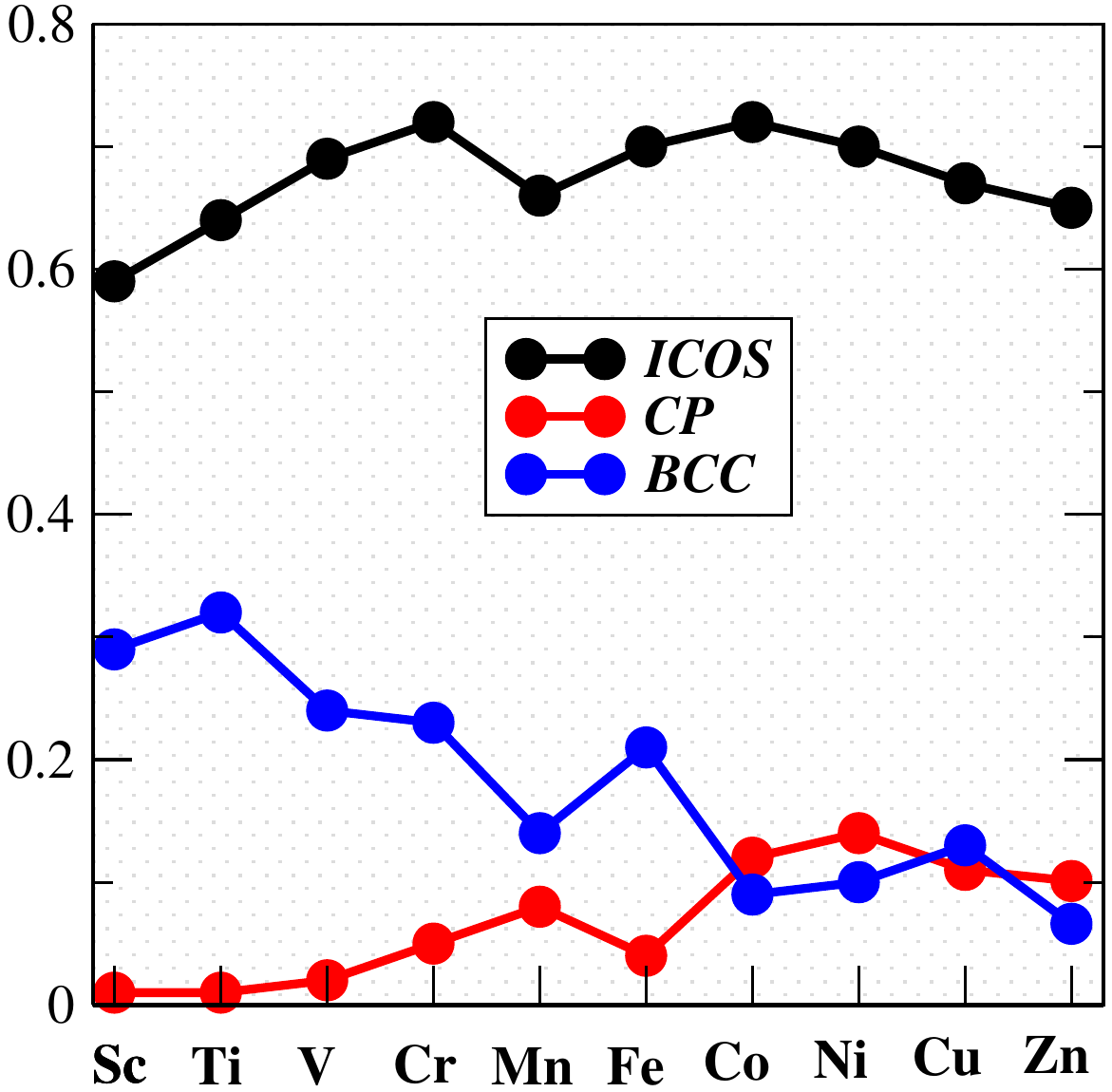}}
\caption{(Colour online) Variation of the most abundant bonded pairs.}
\vspace{.3cm}
\label{fig_Cna}
\end{wrapfigure}

A more detailed three-dimensional description of 
the short range order in these liquid metals  is achieved 
by  applying the 
common neighbour analysis \cite{Andersen} (CNA) method. This 
method allows one to characterize the local environment surrounding each atomic 
pair that contributes to the peaks of the $g(r)$, in terms of 
the number and properties of the 
common nearest neighbours of the pair under consideration.
Each bonded pair of atoms is 
characterized by four indices, with the  
first index being 1 if the pair belongs to the first peak of 
$g(r)$, the second index stands for 
the number of common first neighbours, the  
third index is the number of  
bonds that connect those shared first neighbours and the 
fourth index discriminates among those configurations having the same
first three indices but a different topology. 
The CNA method allows one to discern different local structures 
such as  FCC, HCP, BCC and ICOS. 
For example, the close-packed (CP) structures, FCC and HCP, 
are composed of 142x-type pairs, 
the BCC is typified by 144x and 166x pairs 
whereas the perfect ICOS is characterized by 155x pairs and the 
distorted ICOS is described by the 154x and 143x pairs.

\begin{table}[!t]
\caption{
\label{static}
Calculated values for $r_{\rm min}$ (in \AA),  
coordination numbers CN, $S(q \to 0) $ and isothermal 
compressibilities  $\kappa_T$ (in 10$^{-11}$$m^2$$N^{-1}$ units) 
for the liquid transition metals at 
the thermodynamic states given in table \ref{states}. 
The numbers in 
parenthesis are experimental and/or semiempirical data. }
\vspace{2ex}
\begin{center}
\begin{tabular}{ccccc}
\hline
 $\;\;\;$  & $r_{\rm min}\;\;\;$ & CN $\;\;\;$ & $S(q \to 0)\;\;\;\;$ & 
$\kappa_T\;\;$ \\
\hline
 Sc & 4.27 & 12.8 &   0.020 $\pm$ 0.002 & 2.00 $\pm$ 0.10 $\;\;\;$  (3.25$^a$ )        \\
 Ti & 3.90 & 13.3 & 0.018 $\pm$ 0.001 & 1.28 $\pm$ 0.10 $\;\;\;$   (1.67 $\pm$ 0.02$^b$ )  \\
 V & 3.62 & 12.6 & 0.021 $\pm$ 0.002 & 1.10 $\pm$ 0.10  $\;\;\;$  (1.43$^a$, 1.21$^b$  )   \\
 Cr & 3.47 & 12.5 & 0.020 $\pm$ 0.002 & 0.94 $\pm$ 0.10  $\;\;\;$ (1.00$^a$, 1.10$^b$ )  \\
 Mn & 3.48 & 12.3 & 0.069 $\pm $ 0.003 & 4.9$\pm$ 0.30  $\;\;\;$  (3.00$^a$, 3.74 $\pm$ 0.02)  \\
 Fe & 3.42 & 12.5 & 0.024 $\pm$ 0.002 & 1.24 $\pm$ 0.10 $\;\;\;$ (1.21 $\pm$ 0.02)   \\
 Co & 3.25 & 11.6 & 0.022 $\pm$ 0.002 & 1.10 $\pm$ 0.10 $\;\;\;$ (1.43$^a$, 1.18 $\pm$ 0.02$^b$  )  \\
 Ni & 3.27 & 11.8 & 0.018 $\pm$ 0.002 & 0.95 $\pm$ 0.10 $\;\;\;$ (1.45$^a$,  1.04 $\pm$ 0.02$^b$ ) \\
 Cu & 3.28 & 11.8 & 0.016 $\pm$ 0.002 & 1.10 $\pm$ 0.10  $\;\;\;$  (1.49$^b$ )        \\
 Zn & 3.69 & 12.4 & 0.009 $\pm$ 0.004 & 1.52 $\pm$ 0.60  $\;\;\;$  (2.34$^b$ )        \\
\hline
\multicolumn{5}{l}{
$^a$  \cite{Marcus},
$^b$  \cite{Blairs}}
\end{tabular}
\end{center}
\vspace{-6mm}
\end{table}

The CNA study is performed on inherent structures in which 
the ions are brought to local minima of the potential 
energy surface. 
For each metal, this calculation was 
made for four/five ionic configurations, 
well separated in time ($\approx 15.0 {-} 20.0$ ps), and the 
results were averaged.  These are plotted in 
figure \ref{fig_Cna} where it is noticed that 
the five-fold symmetry dominates, as 
the sum of perfect and distorted  
ICOS structures varies between  $\approx$ 59\% (l-Sc) and 
$\approx$ 72\% (l-Cr, l-Co) of the pairs. 
The amount of local BCC-type pairs is also remarkable as it ranges from 
$\approx$ 32\% (l-Ti) to  
$\approx$ 6\% (l-Zn) of the pairs. 
Finally, the CP-type pairs are almost absent in the early transition metals
but become significant for Mn and for the late elements (Co to Zn).
This behaviour of the CP vs BCC local structures correlates with the
corresponding phase diagrams of the elements, since Sc to Fe melt
from a BCC phase, while Co to Zn do so from CP structures, either
FCC (Co, Ni, Cu) or HCP (Zn). The phase diagram of Mn is rather complex, and
the high temperature BCC phase only exists down to 100 K below the melting point,
where it transforms into FCC, and this latter phase appears to somehow partially 
survive locally also after melting.

\subsection{Dynamic properties}

We  evaluated several dynamic magnitudes and their 
associated time correlation functions. Due to the  
macroscopically isotropic behaviour of the fluid, those 
correlation functions with a 
$\vec{q}$-dependence are transformed into 
$|\vec{q}|$-dependent magnitudes.

\subsubsection{Single-particle dynamics}
 
We report results for the 
self-diffusion coefficient, $D$, which, along with the shear viscosity,  
plays an important role in the study of crystal nucleation and 
growth in  metallic melts. We evaluated 
the velocity autocorrelation function, $Z(t)$, and  the mean square 
displacement, $\delta R^2(t)$, 
of a tagged ion in the fluid, for all the $3d$ liquid metals.
The diffusion coefficient can be determined through both
of these functions, and
both approaches  yielded practically the same result for 
$D$, which is given in  
table \ref{dynamic} along with 
the available experimental/semiempirical data.   

\begin{table}[htb]
\caption{
\label{dynamic}
Calculated values of the 
self-diffusion coefficient, $D$ (\AA$^2$/ps), 
 adiabatic sound velocity $c_\text{s}$ (m/s) and  
shear viscosity $\eta$ (GPa$\,\cdot\,$ps) 
for the liquid transition metals at 
the thermodynamic states given in table~\ref{states}. The numbers in 
parenthesis are experimental and/or semiempirical data. }
\vspace{2ex}
\begin{center}
\begin{tabular}{cccc}
\hline
 $\;\;\;$  & $D\;\;\;$ &  $c_\text{s}\;\;\;$ & $\eta\;\;\;\;$ \\ 
\hline
 Sc $\;$ & 0.55 $\pm$ 0.01 &  4400 $\pm$ 200 & $\;$ 2.10 $\pm$ 0.10  \\
 Ti $\;$ & 0.49 $\pm$ 0.01 (0.53 $\pm$ 0.02$^a$) &  4640 $\pm$ 100 (4407 $\pm$ 108$^b$) & $\;$ 2.85 $\pm$ 0.15  (2.20$^c$)  \\
 V $\;$  & 0.51 $\pm$ 0.01 & 4725 $\pm$ 200 (4742$^b$) & $\;$ 3.30 $\pm$ 0.20  \\
 Cr $\;$ & 0.47 $\pm$ 0.01 & 4500 $\pm$ 200 (4298$^b$ )  & $\;$ 3.70 $\pm$ 0.20     \\
 Mn $\;$ & 0.46  $\pm$ 0.02 &  3100 $\pm$ 150 (3381$^d$)  & $\;$ 2.50 $\pm$ 0.20     \\ 
 Fe $\;$ & 0.37 $\pm$ 0.02 (0.355$^e$) & 3950 $\pm$ 150 (3820 $\pm$ 150$^f$) & $\;$ 5.00 $\pm$ 0.30 (5.30$^g$)  \\
 Co $\;$ & 0.60 $\pm$ 0.01 & 3700 $\pm$ 150 (4048$^d$) & $\;$ 3.20 $\pm$ 0.20 (4.12$^h$, 3.60$^i$)    \\ 
 Ni $\;$ & 0.38 $\pm$ 0.01 (0.380 $\pm$ 0.006$^{j}$) & 4250 $\pm$ 150  (4040 $\pm$ 150$^g$) & $\;$ 4.10 $\pm$ 0.20 (4.37$^h$)  \\
 Cu $\;$ & 0.27 $\pm$ 0.01  (0.37 $\pm$ 0.006$^{l}$) & 3550 $\pm$ 150 (3481 $^d$)  & $\;$ 4.80 $\pm$ 0.20 (4.38$^m$)  \\
 Zn $\;$ & 0.23 $\pm$ 0.01  (0.24 $\pm$ 0.006$^{n}$) & 3150 $\pm$ 250 (2849 $^d$)  & $\;$ 2.40 $\pm$ 0.30 (3.50$^{e,p} $)  \\
\hline
\multicolumn{4}{l}{
$^a$   \cite{SaidHorbatch},   
$^b$   \cite{Casas},  
$^c$   \cite{Agaev}, 
$^d$   \cite{Blairs}, 
$^e$   \cite{IidaBook}, 
$^f$  \cite{Nash}, 
$^g$  \cite{ShimoBook},  
$^h$  \cite{Assael}, 
$^i$  \cite{Egry},  
$^j$  \cite{Chato},   
$^l$   \cite{Meyer}, 
$^m$   \cite{Hirai}, 
$^n$  \cite{Nachtrieb},
$^p$  \cite{Assael2} }
\end{tabular}
\end{center}
\vspace{-5mm}
\end{table}

\subsubsection{Collective dynamics}

In a liquid, the dynamics of 
density fluctuations is usually described by the 
intermediate scattering function, $F(q, t)$, which is defined as 
autocorrelation function of the microscopic 
wavevector-dependent density \cite{Balubook}. 
The Fourier transform (FT) of $F(q, t)$ into the frequency 
domain leads to the dynamic structure 
factor, $S(q,\omega)$, which can be directly measured by either 
INS and/or IXS experiments.

Another important collective magnitude is the current 
due to the overall motion of the particles. It is a  
vectorial magnitude which is usually split into 
its longitudinal and transverse components with respect to $\vec{q}$. Then, 
the longitudinal, $J_\text{L}(q, t)$, and transverse $J_\text{T}(q, t)$, 
current correlation functions are
obtained as autocorrelation functions of the corresponding quantities \cite{Balubook}. 
Their time FT yields the associated spectra, $J_\text{L}(q, \omega)$  and  $J_\text{T}(q, \omega)$, 
respectively.

\begin{figure}[!t]
\centerline{\includegraphics[width=0.95\textwidth,clip]{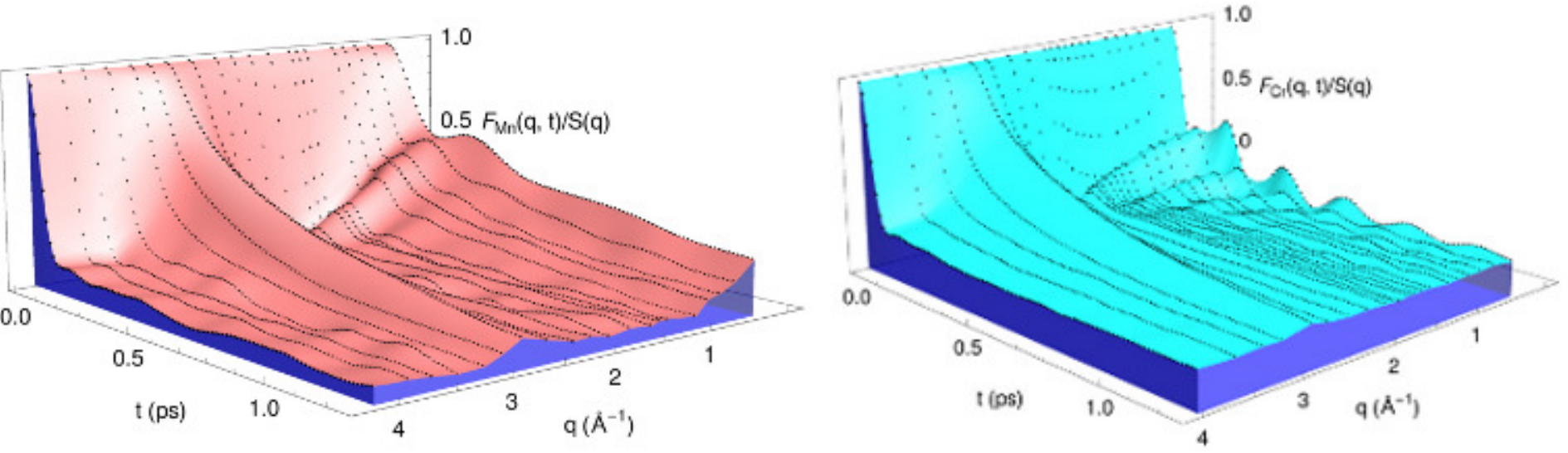}}
\caption{(Colour online) Intermediate scattering function, $F(q,t)/S(q)$, of 
l-Mn at $T=1550$ K and l-Cr at $T=2173$ K for 
several $q$ values. } 
\label{FktMnCr}
\end{figure}
\begin{figure}[!t]
\centerline{\includegraphics[width=0.975\textwidth,clip]{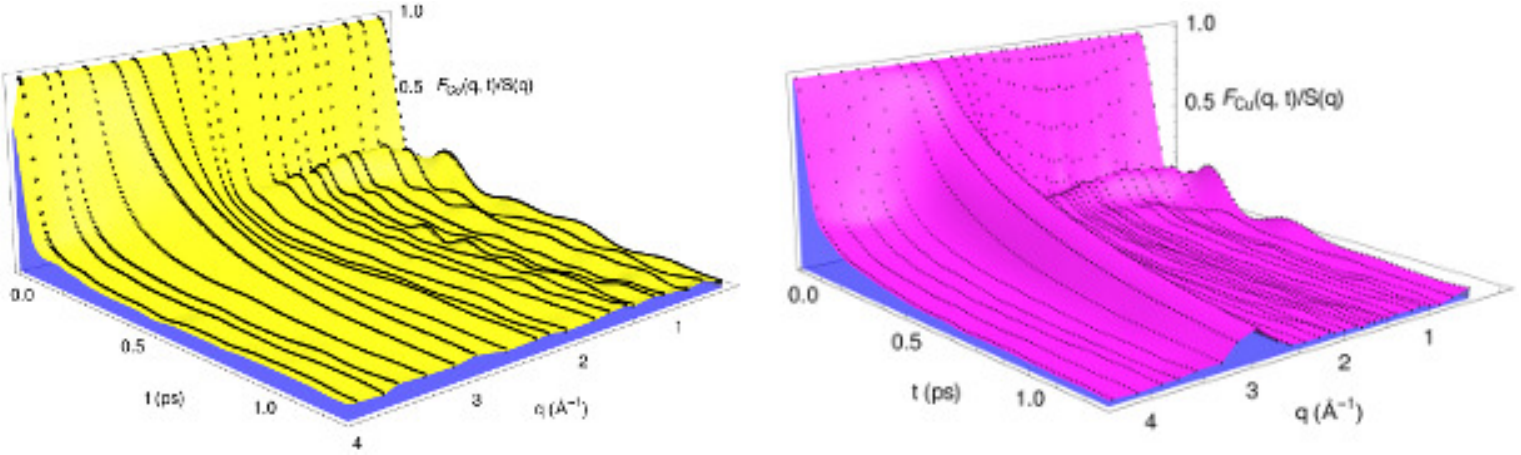}}
\caption{(Colour online) Same as the previous figure but for 
	l-Co at $T=1850$ K and l-Cu at $T=1423$ K. } 
\label{FktCoCu}
\end{figure}
\begin{figure}[!b]
\centerline{\includegraphics[width=0.95\textwidth,clip]{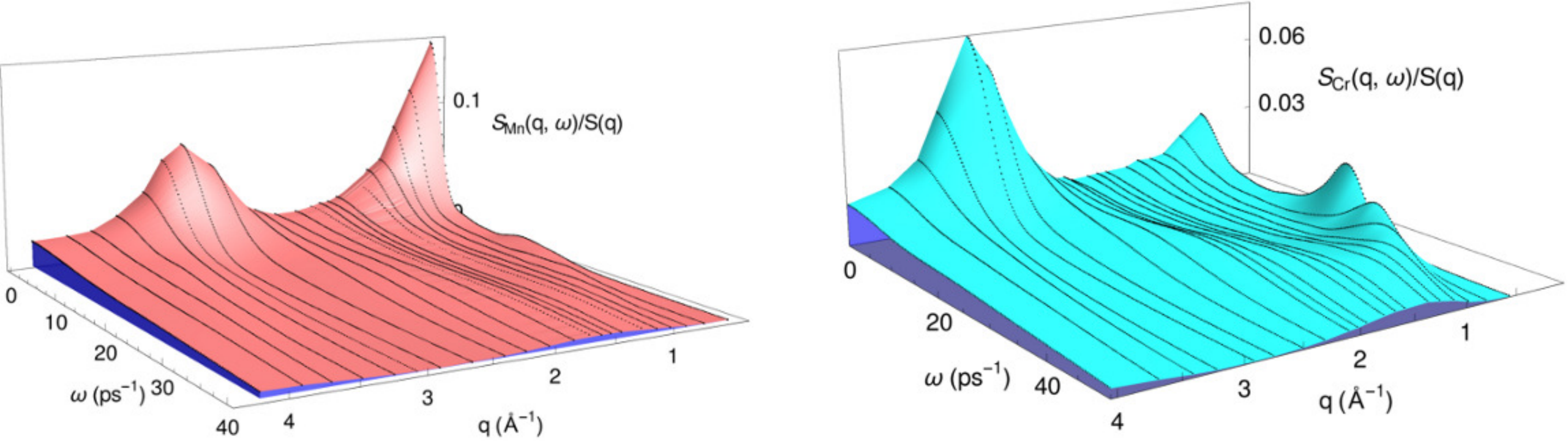}}
\caption{(Colour online) Dynamic structure factor, $S(q,\omega)/S(q)$, of 
l-Mn at $T=1550$ K and l-Cr at $T=2173$ K for 
several $q$ values. } 
\label{SqwMnCr}
\end{figure}

In figures  \ref{FktMnCr}--\ref{FktCoCu}, we  plotted,  
for some metals and for some range of $q$-values, the obtained 
AIMD results for the $F(q, t)$. 
For small $q$-values, the $F(q,  t)$ show an oscillatory behaviour,  
indicative of wave propagation, which is superposed on another decaying 
component, indicative of relaxation modes. 
The oscillations become weaker with 
increasing $q$-values and disappear at around $q \approx 4/5 q_p$,
where $q_p$ denotes the position of the main peak of $S(q)$.
For larger $q$ values, the 
decaying component becomes dominant leading at $q \approx q_p$ to a very 
slow monotonous decay of the $F(q, t)$. 
The relative strength of both 
components depends on the system and we observe that l-Mn has 
the weaker oscillations superposed on a very strong 
diffusive component.

\begin{figure}[!t]
\centerline{\includegraphics[width=0.95\textwidth,clip]{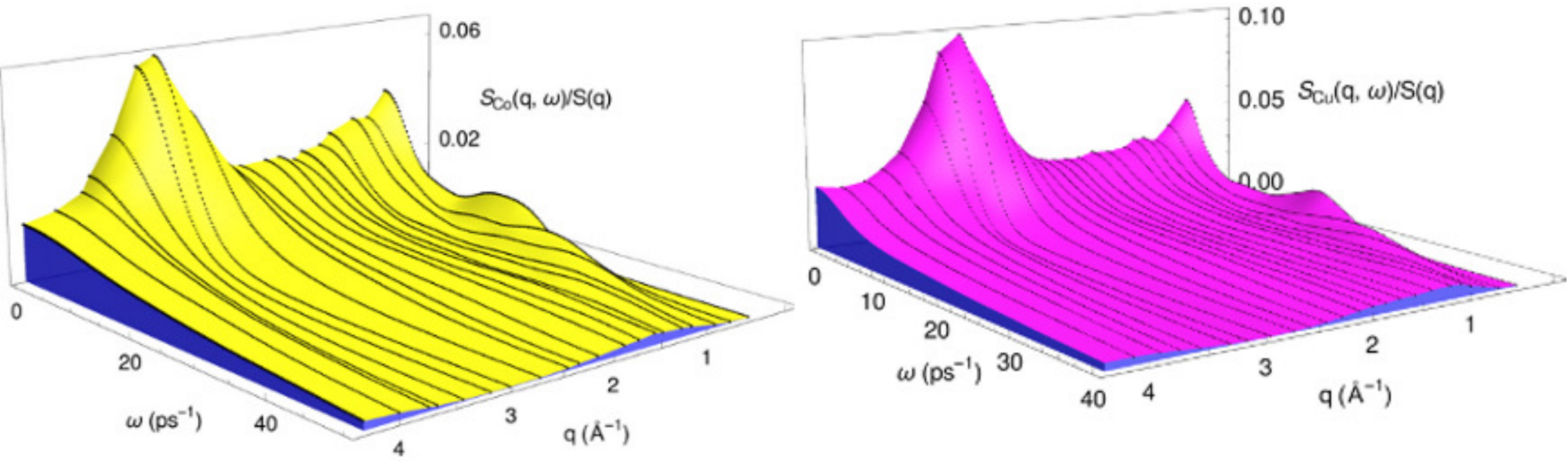}}
\caption{(Colour online) Same as the previous figure but for 
	l-Co at $T=1850$ K and l-Cu at $T=1423$ K. } 
\label{SqwCoCu}
\end{figure}

\begin{figure}[!t]
\centerline{\includegraphics[width=0.94\textwidth,clip]{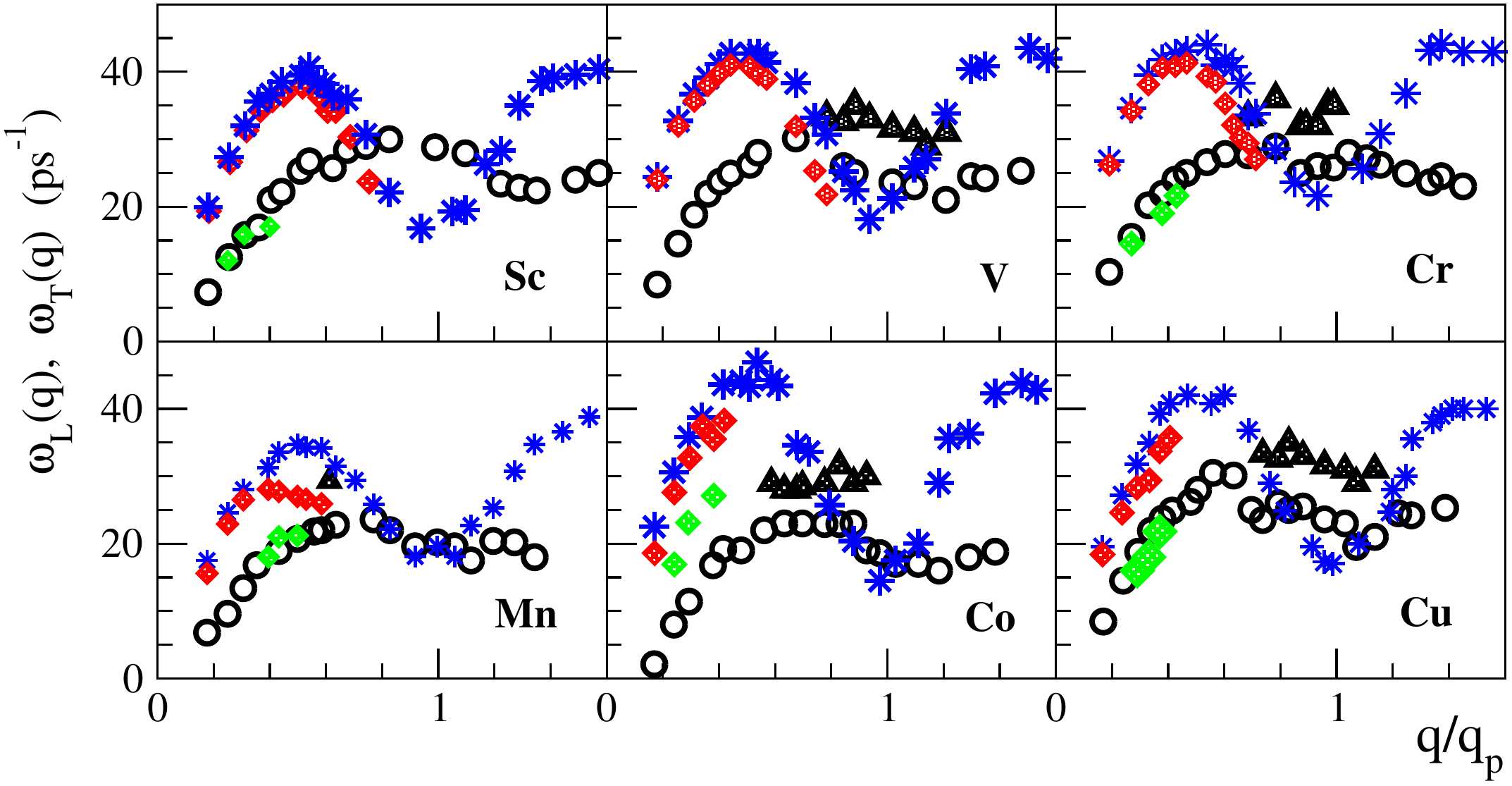}}
\vspace{-2mm}
\caption{(Colour online) Dispersion relations for several 3$d$ liquid metals. 
Red diamonds and stars: longitudinal dispersion obtained from 
the AIMD results for the positions of the inelastic peaks
in the $S(q,\omega)$ and from the maxima in the spectra of the longitudinal
current, $J_\text{L}(q, \omega)$, respectively.
Open circles and triangles: transverse dispersion from the positions
of the peaks in the spectra $J_\text{T}(q, \omega)$.
The green lozenges  
are the positions of the TA excitation modes found in the calculated 
dynamic structure factors, $S(q, \omega)$. 
\label{DisperLONTRANSA}
}
\end{figure}

Figures \ref{SqwMnCr}--\ref{SqwCoCu} show the  
associated $S(q, \omega)$. 
The behaviour of the calculated 
$S(q, \omega)$ is qualitatively similar for all 
the 3$d$ metals studied in this paper. 
Specifically, the  
$S(q, \omega)$ show side-peaks up 
to $q$-values $\approx (3/5)q_p$, which thereafter evolve into shoulders 
lasting up to $q \approx (4/5)q_p$.  Therefrom, the  
$S(q, \omega)$  show a monotonous decreasing behaviour.
Notice that for small $q$-values, the $S(q, \omega)$ of l-Mn takes 
relatively large values when $\omega \to 0$; this is due to the 
important diffusive component of its $F(q, t)$ at 
those small  $q$-values. 
From the positions of the side-peaks, $\omega_m(q)$,  
in the $S(q, \omega)$, a dispersion relation for the density 
excitations was obtained and  is plotted 
in figure \ref{DisperLONTRANSA}. 
In the hydrodynamic region (small $q$), the slope of 
the dispersion  relation  curve is the  $q$-dependent adiabatic sound  
velocity $c_\text{s}(q)=v_\text{th}\sqrt{\gamma/S(q)}$, with $v_\text{th}=\sqrt{k_{\text B}T/m}$ being 
the thermal velocity and $\gamma$ being the ratio of specific heats. 
In the $q\rightarrow 0$ limit, the $c_\text{s}(q)$ reduces to the bulk 
adiabatic sound velocity $c_\text{s}$. 
However, the small size of the simulation box implies that the
smallest attainable $q$ value, namely $q_{\rm min}$, is not 
small enough so as to permit an accurate determination of  $c_\text{s}$; 
nevertheless, an approximate estimation can be obtained from 
value of $\omega_m(q_{\rm min})/q_{\rm min}$. From the values  
$q_{\rm min}$= 0.43~\AA$^{-1}$ (Sc), 0.506~\AA$^{-1}$ (Ti), 0.508~\AA$^{-1}$ (V), 0.561~\AA$^{-1}$ (Cr),
0.513~\AA$^{-1}$ (Mn), 0.570~\AA$^{-1}$~(Fe), 0.525~\AA$^{-1}$ (Co), 0.547~\AA$^{-1}$ (Ni),  0.500~\AA$^{-1}$ (Cu), and
0.553~\AA$^{-1}$ (Zn), we  obtained 
the values given in table \ref{dynamic} where they are 
compared with the available experimental/semiempirical data.

We  also evaluated the 
longitudinal and transverse currents, $J_\text{L}(q,t)$ and $J_\text{T}(q,t)$, as well as 
their respective spectra, $J_\text{L}(q, \omega)$ and $J_\text{T}(q, \omega)$. 
The obtained $J_\text{L}(q, \omega)$ show one peak for each $q$-value 
and  the frequencies associated to those peaks, $\omega_\text{L}(q)$, 
constitute the longitudinal dispersion relation for the associated 
collective modes. 
 Figure \ref{DisperLONTRANSA}  represents the obtained dispersion relations, 
 which show the 
typical behaviour found in other liquid systems \cite{Balubook}.

\begin{figure}[!t]
\centerline{\includegraphics[width=0.95\textwidth,clip]{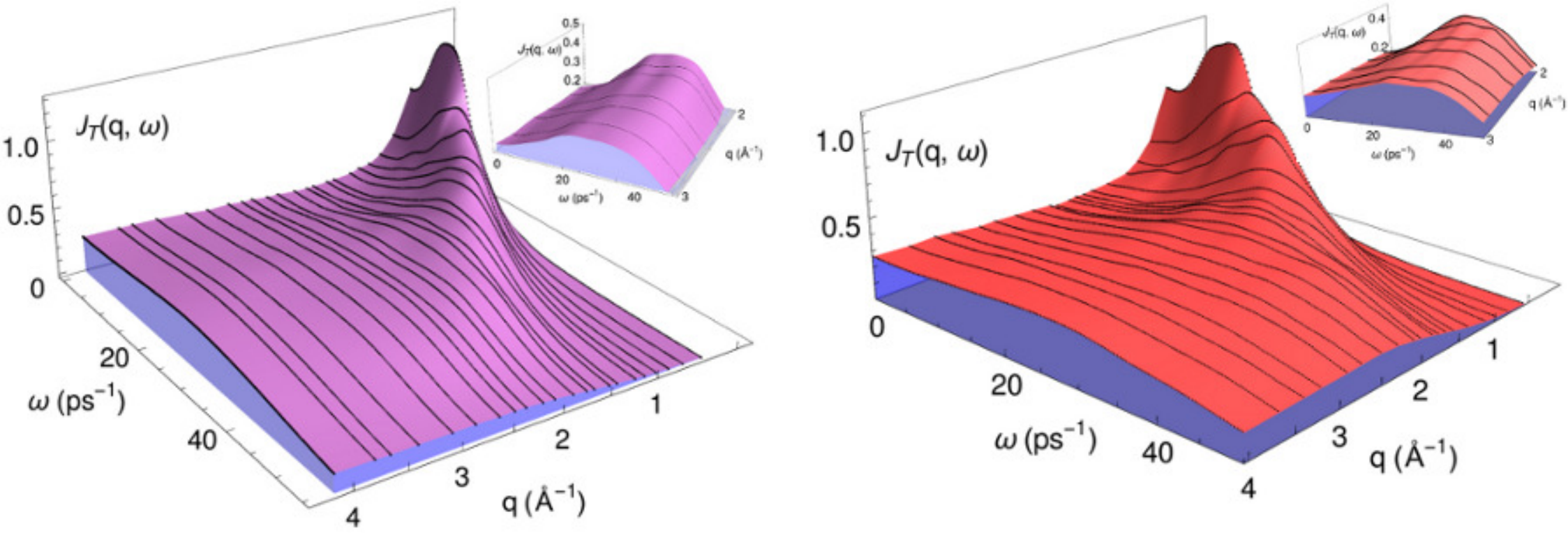}}
\caption {(Colour online) Transverse current spectra, $J_\text{T}(q, \omega)$, 
of l-V at $T=2173$ K and l-Cr at $T=2173$~K at 
several $q$-values. The inset gives a detailed view 
of the region where the $J_\text{T}(q, \omega)$ displays two maxima. }
\label{CtkwVaCr}
\end{figure}

As for the  $J_\text{T}(q, \omega)$, it 
may exhibit peaks within some $q$-range, 
which are related to propagating shear waves. 
For all the metals studied in this 
paper,  the associated $J_\text{T}(q=q_{\rm min}, \omega)$  already showed a peak  and 
with increasing $q$-values the associated frequency of the peak, $\omega_\text{T}(q)$ also increased and reached 
a maximum value  for $q \approx (2/3)q_p$; therefrom it  decreased and  
 vanished  at $\approx 3.0 q_p$. 
An example of this behaviour 
is given in 
figure \ref{CtkwVaCr} which shows the calculated $J_\text{T}(q, \omega)$ for 
l-V and  l-Cr, as a function of $\omega$, and for 
a range 0 $< q \leqslant 4.0$~\AA$^{-1}$.   

Moreover, for some 
metals (l-V, l-Cr, l-Co, l-Cu, l-Ni and l-Zn)  we found that 
the $J_\text{T}(q, \omega)$ exhibited, within 
the $q$-region 0.8 $\leqslant q/q_p \leqslant 1.2$,  
another peak with a higher frequency. 
Figure \ref{CtkwVaCr} shows, for l-V and l-Cr,  a closer view of 
that specific region where the two maxima are visible. 
For the other metals, (i.e., l-Sc, 
l-Ti, l-Mn and l-Fe) the results are not conclusive because, besides 
the lower frequency peak,  we 
also found higher frequency shoulders instead of peaks. 

The appearance of another, high frequency branch in the transvese dispersion relation, had 
been reported first in some liquid metals 
at high pressure, i.e., Li, Na, Fe, Al and Pb, \cite{MGGFe,BRSS,BDJ,BDJ1}. 
Subsequently, it was also  found in 
l-Tl and l-Pb \cite{Taras18,BDJ,BDJ1} at ambient pressure. 

From the calculated $J_\text{T}(q, t)$, we  also  
estimated 
\cite{Balubook} the shear viscosity 
coefficient $\eta$, as follows.  
The memory function representation of $J_\text{T}(q,t)$, 
namely
\begin{equation}
\tilde{J_\text{T}}(q,z)=\frac{k_{\text B} T}{m}
	\left[  z+\frac{q^2}{\rho m}\tilde{\eta}(q,z)\right]^{-1}, 
\end{equation}
where the tilde denotes the Laplace transform, introduces a generalized shear 
viscosity coefficient  $\tilde{\eta}(q,z)$. 
The area under the normalized $J_\text{T}(q,t)$ gives $m \tilde{J_\text{T}}(q,z=0)/(k_{\text B} T)$, 
from which $\tilde{\eta}(q,z=0)\equiv\tilde{\eta}(q)$ can be obtained. 
Extrapolation to $q\rightarrow 0$ 
gives the standard shear viscosity coefficient $\eta$.  
Different extrapolation functions lead to different values of $\eta$,
which are portrayed in the uncertainty assigned to the results.
Table~\ref{dynamic} shows 
these results along 
with the available experimental/semiempirical data.

Recently, it was suggested \cite{HosoTGa} that 
transverse-like low-energy excitations might 
be observed as weak  
shoulders located in the region between the quasielastic and the 
longitudinal inelastic peaks 
of the $S(q, \omega)$. 
These excitations  are usually visible within a  
small $q$-range around $q_p/2$,  because for smaller/greater 
$q$-values they  are overcome by 
the quasielastic/inelastic peaks.
These transverse-like excitations were first 
detected in the IXS spectra of l-Ga \cite{HosoTGa} at 313 K and subsequently 
in l-Cu, l-Sn, l-Na, l-Fe, l-Ni     
and l-Zn \cite{RGGAgNi,HosoTCuFe,HosoTSn,HosoZana,HosoZana1,MonacoTNa}.   
\begin{figure}[!t]
\centerline{\includegraphics[width=0.825\textwidth,clip]{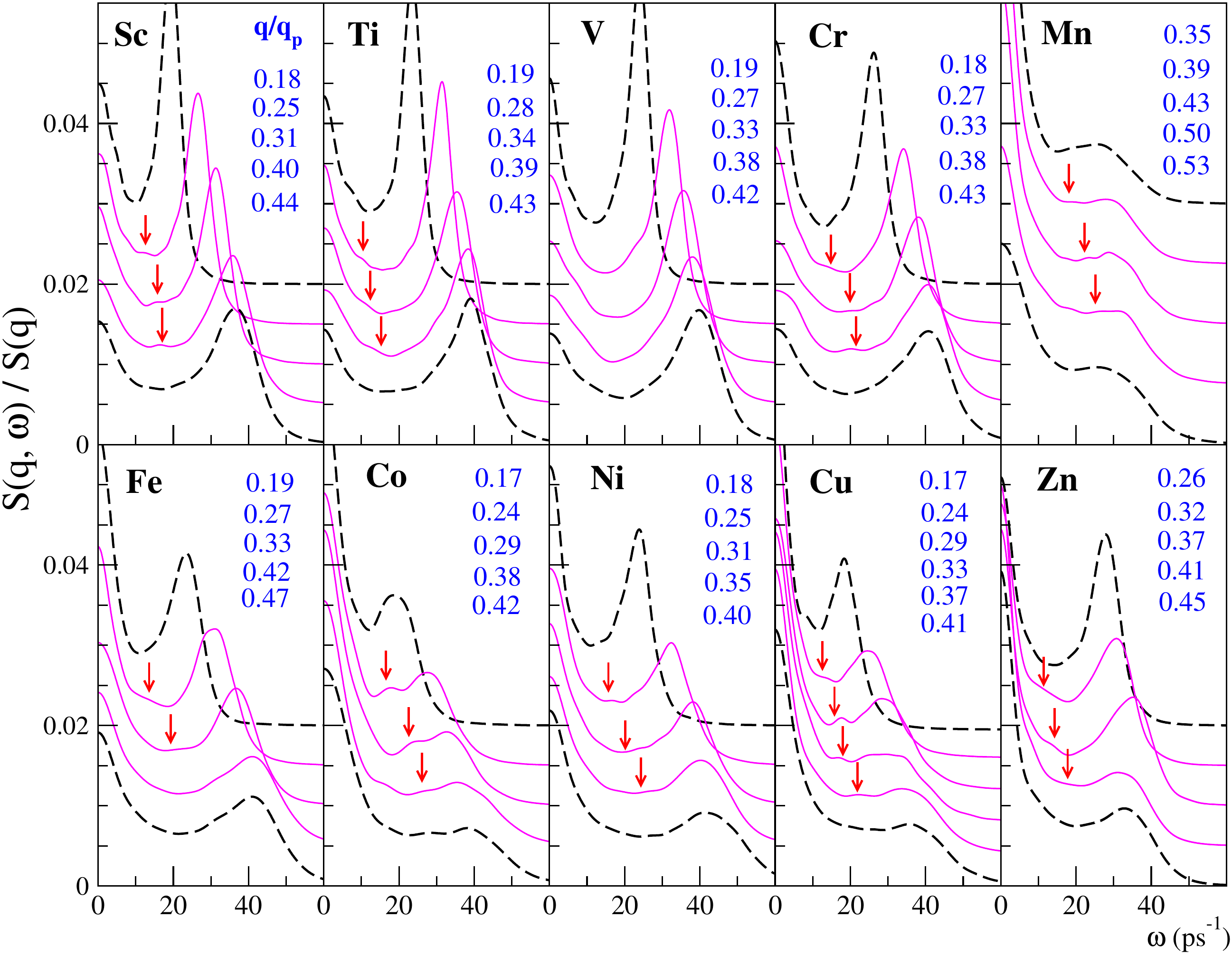}}
\caption{(Colour online) AIMD calculated dynamic structure factors $S(q,\omega)/S(q)$ 
of the 3$d$ liquid transition metals. 
They are plotted for different values 
of  $q/q_p$ (top to bottom). 
The vertical scales are offset for clarity. The 
arrows point to the locations of the   
transverse-like excitations.  
\label{DisperTRANS2a}
}
\end{figure}
We   analyzed the results 
for $S(q, \omega)$ and found that  
within the range 1.00 \AA$^{-1}$ $\leqslant q \leqslant$ 1.45 \AA$^{-1}$, 
some weak shoulders appear in the $\omega$-region located 
between the quasielastic and the inelastic peaks. This is shown 
in figure \ref{DisperTRANS2a}; moreover,  
the energies associated to these shoulders are close to those 
corresponding to the 
peaks in the transverse current spectra. 
The real physical origin of these excitations is still a
matter of debate, and some 
 alternative interpretations, for instance heat
waves, were proposed within the framework of the GCM approach \cite{Bryk-Na}.

\subsection{Electronic properties: density of states}

The partial and total electronic density of
states, $n(E)$, were obtained from the
self-consistently determined eigenvalues; they
were averaged over four/six ionic configurations
well separated in time ($\approx 15.0$~ps),
 and the sampling 
of the Brillouin zone was performed 
using an $8\times8\times8$ Monkhorst-Pack set.
\begin{figure}[!t]
	\centerline{\includegraphics[width=0.895\textwidth,clip]{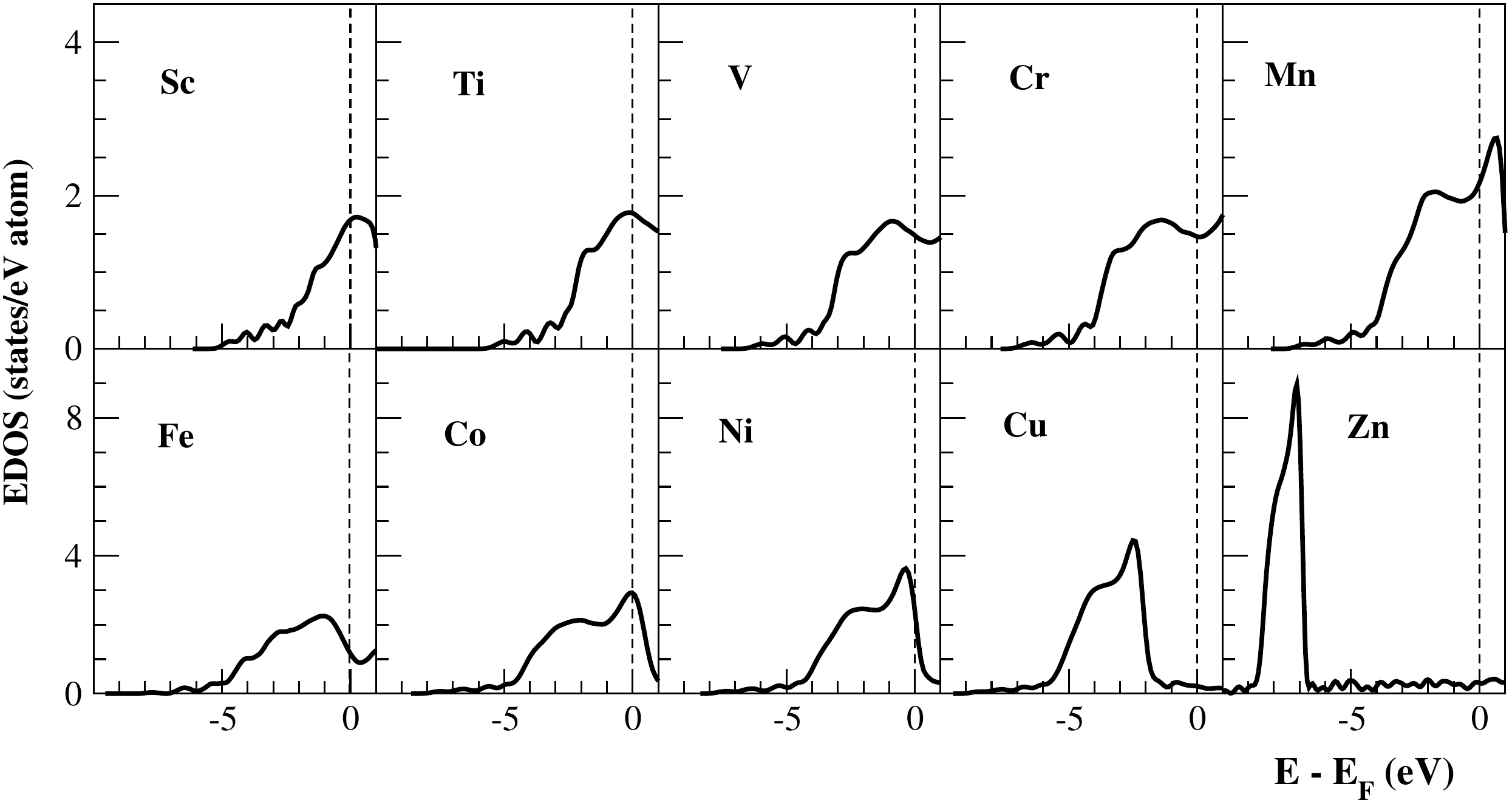}}
	\caption{Total electronic density of states. 
		Notice the different scale in the upper and lower panels.
		\label{dosfig}
	}
\end{figure}

Figure \ref{dosfig} shows the obtained results for the
electronic 
total $n(E)$ associated to the 
outer valence electrons, which  
is dominated by the 3$d$ states, 
except in the early transition elements, where the
$d$ band occupation is low. 
Notice that for the metals Sc to Mn we are not showing the
contribution from the $3s$ and $3p$ atomic states, which were 
included in the self-consistent calculation but lie way below
the
 Fermi energy.  
In the figure we observe the progresive filling of the $d$ band,
which is complete at Cu, together with a decrease of its width
and an increase of its height as the occupation rises.

\section{Conclusions}
\label{conclusions}

An {\it ab initio} molecular dynamics simulation 
method was used to calculate several  
static and dynamic properties of the 
bulk liquid $3d$ transition metals 
 near their respective melting points.  

The results for the static structure, as characterized by  
the static structure factor $S(q)$, 
show a  good agreement with the available experimental data. 
The only exception is l-Sc where a clear difference is observed
in the phase of the oscillations, but this may be due to
inaccuracies in the old experimental data, similar to the
case of the old data for l-Ti.
The  $S(q)$ display, in most cases, an asymmetric
shape in the 
second peak which qualitatively agrees
with the recent ND and/or XD data. This feature is  
related to the appearance of icosahedral short-range order in the liquid metal. 
This  is confirmed by a more detailed study of 
the liquid static structure by using the CNA method,
which  revealed a marked abundance of five-fold structures.
On the other hand, secondary local structures also present in the liquid correlate
rather well with the solid phase from where the metals melt.

The calculated dynamic structure 
factors, $S(q,\omega)$, show side-peaks which are 
indicative of collective density excitations. 
A detailed analysis of the $S(q, \omega)$  revealed 
some type of excitations which have similar features as 
the transverse-like excitation modes  
found in IXS and INS data for 
several liquid metals near their 
respective melting points. 

From the calculated 
longitudinal and  transverse current correlation functions, we  evaluated 
the associated spectral functions and the corresponding dispersion relations. 
For some metals, we  obtained two branches of transverse 
modes, with the high frequency branch appearing over a limited $q$-range 
which is always located in the second pseudo-Brillouin zone. 

Results are also  reported for several transport coefficients  
such as the self-diffusion, adiabatic sound velocity and shear viscosity.  
Although the calculated values for the adiabatic sound velocity  
show a reasonable agreement with the available 
experimental data, the lack of experimental data for the self-diffusion and the 
shear viscosity provides a further relevance of these results. 

\section*{Acknowledgements}

This work has been supported by Junta de Castilla y Leon (project VA124G18).
LEG and DJG also acknowledge the funding from the Spanish Ministry 
of Economy and Competitiveness (Project PGC2018-093745-B-I00), which is
partially supported by the EU FEDER program.

\ukrainianpart

\title{Визначення з перших принципів деяких статичних і динамічних властивостей  3$d$
	рідких перехідних металів поблизу плавлення
}
\author{Б.Г. дель Ріo,  К. Паскуаль, О. Родрігес, Л.Є. Гонзалес, Д.Дж. Гонзалес}
\address{Відділ теоретичної фізики, університет  Вальядоліда, 47011 Вальядолід, Іспанія
}

\makeukrtitle 

\begin{abstract}
	Ми повідомляємо  \textit{an initio} дослідження симуляціями  молекулярної динаміки деяких статичних і динамічних властивостей  3\textit{d} рідких перехідних металів. Обчислені  статичні структурні фактори показують якісне узгодження з наявними експериментальними даними, а їх другий пік має асиметричну форму, яка передбачає, що існує значний локальний  ікосаедричний порядок. Динамічна структура  виявляє розповсюдження  флуктуацій густини, дисперсійне співвідношення яких було обчислено;   крім того їх довгохвильова границя є
	сумісною з відповідною їм експериментальною швидкістю звуку.  Представлено результати для спектральних функцій і для відповідних дисперсійних співвідношень. Ми також аналізуємо можливу появу поперечно-подібних низькоенергетичних збуджень в обчислених динамічних структурних факторах. Обчислено декілька коефіцієнтів переносу і порівняно з наявними експериментальними даними. 
	
	\keywords рідкі метали, першопринципні розрахунки 
\end{abstract}

\lastpage
\end{document}